# AR-Therapist: Design and Simulation of an AR-Game Environment as a CBT for Patients with ADHD

**Saad Alqithami\*** , **Musaad Alzahrani, Abdulkareem Alzahrani and Ahmed Mustafa**

Department of Computer Science, Albaha University; Albaha 65799, Saudi Arabia; malzahr@bu.edu.sa (M.A.), ao.alzahrani@bu.edu.sa (A.A.), amyosof@bu.edu.sa (A.M.)
\* Correspondence: salqithami@bu.edu.sa (S.A.)

**Abstract:** Attention Deficit Hyperactivity Disorder is one of the most common neurodevelopmental disorders in which patients have difficulties related to inattention, hyperactivity, and impulsivity. Those patients are in need of a psychological therapy use Cognitive Behavioral Therapy (CBT) to enhance the way they think and behave. This type of therapy is mostly common in treating patients with anxiety and depression but also is useful in treating autism, obsessive compulsive disorder and post-traumatic stress disorder. A major limitation of traditional CBT is that therapists may face difficulty in optimizing patients' neuropsychological stimulus following a specified treatment plan. Other limitations include availability, accessibility and level-of-experience of the therapists. Hence, this paper aims to design and simulate a generic cognitive model that can be used as an appropriate alternative treatment to traditional CBT, we term as "AR-Therapist." This model takes advantage of the current developments of augmented reality to engage patients in both real and virtual game-based environments.

## 1. Introduction

Attention Deficit-Hyperactivity Disorder (ADHD) has been an increasing concern in the past few decades. It has undefined etiology, as a heterogeneous developmental disorder, leading to bias- and extensive-diagnostic evaluations when examining patients through traditional clinical interviews and ratings of patients' behaviors [2,3]. ADHD in underage patients can be observed in patients' hyperactivity (i.e., inability to control their impulses) and difficulty to pay attention which, as a result, will minimally intervene with their social engagements and the continues progress of their normal daily lives. In adulthood, patients with ADHD may have trouble managing time, being organized, setting goals, and holding down a job, which may lead to other problems related to relationships, self-esteem, and possibly addiction. The treatment of many psychological disorders, such as ADHD, can be through a well-known type of psychotherapy called Cognitive Behavioral Therapy (CBT). CBT involves patients in multiple psychosocial interventions in order to improve their mental health. This treatment requires patients to go through multiple sessions with specialized therapists. In a case of ADHD, those sessions can be of increasing order of difficulty/complexity to provide patients with the ability to expand their cognitive capabilities and to overcome their current behavioral limitations.

This highlights the usefulness of utilizing intelligent and immersive technologies, e.g., Augmented Reality (AR) and Virtual Reality (VR), for their promising results that have been stated in previous studies [4–9]. As the name suggests, augmented reality is the technology of combining real and 3D rendered virtual contents through a real-time interactive environment [10]. Whereas, the technology of virtual reality replaces real world with a computer-generated graphics via head mounted display [11]. In other words, the user in VR environment totally isolates the player from the real world while the AR optimizes the interactions with 3D objects in a real-world environment [12]. A major benefit of employing such technologies (i.e., VR and AR) to mimic traditional CBT besides the multi-level of virtuality is the automated capturing of a multi-sensory data, e.g., measurements of head movement and the angle of vision are recorded. These recordings help in providing quantitative support for the diagnosis, care and treatment of ADHD which have the ability to replace current labor-intensive techniques involving paper-and-pencil methods or manual video analysis when collecting data. Unfortunately, previous proposed solutions fail to (a) overcome language and cultural barriers for diverse patients and to (b) employ the power of augmented reality by rendering 3D objects and avatars rather than solid textual instructions, which are essentials to increase patients' engagements and to speed-up the recovery time.

The belief is that there is no statistically significant difference between the ADHD patients who will be treated using traditional CBT and those who are treated by an automated-/online-system employing current advancements of augmented reality; although, online CBT may exceed traditional methods by accelerating recovery time and saving money and resources for both government and patients. This is due to achieving a concept of "a therapist for each patient" as the system mimics the therapist roles through augmented reality techniques that provide it with features including: adaptiveness, smartness, responsiveness, and accuracy. Other advantages are availability, accessibility, and assurance of the therapist's level-of-experience which cannot be guaranteed in traditional CBT. Therefore, this paper extends the work presented by the authors in [1] by introducing an AR-based system through modeling and simulation called "AR-Therapist" that serves as an online CBT. We design an augmented reality game using both Microsoft-HoloLens emulator—to tests mixed reality apps on a PC without a physical HoloLens— and Unity —a cross-platform game engine— applications to design a testbed for the system. Further, we defined set of measurements to evaluate the proposed system based on the idea of an increase in patient correct attention to choose a predefined object contributes positively to their performance index which means they are following along with their treatment plan.

The article is organized as follows: Section 2 provides an overview of some of the related work on the different techniques used as a CBT through multiple virtuality techniques and to whether such treatments support our claim. Section 3 proposes a formal modeling to main parameters used to define the conceptual model, and we quantify and propose measurements to identify which parameters contributes most to the increase of the patient's recovery performance and their applicability to the treatment plan. Section 4 contributes an extensive experiment to support the proposed measurements by implementing a simple game that reflects the usefulness of traditional therapists. Section 5 details a hypothetical case study to contribute to the validity of our model and to provide detailed discussion. Section 6 draws conclusions and points to future possibilities.

## 2. Literature Review

Attention Deficit-Hyperactivity Disorder has undefined etiology as a heterogeneous developmental disorder to involve hyperactivity and distractibility as well as difficulties with constant attention, impulsive control disorder and impaired cognitive flexibility, especially in problem solving and behavioral management [13,14]. Many studies have indicated the potential benefits of Virtual Reality (VR) and Augmented Reality (AR) exposure therapy for many types of mental disorders [4–9]. In a study by Parsons, et al. [4], attention performance was compared between 10 children with ADHD and 10 normal control children in a VR classroom. The results showed that children with ADHD are more impacted by distraction in the VR classroom. In spite of that, Ben-Moussa, et al. [15] proposed a conceptual design of an exposure therapy system called DJINNI for social anxiety disorder. DJINNI integrates the AR and VR technologies to provide more effective exposure therapy solutions for patient with social anxiety disorder.

Cho, et al. [16] conducted a study that examined usefulness to use VR for rehabilitation. Their study involved 30-participants with ADHD. The participants were trained in one of three groups: a VR group with a head-mounted display, a desktop equivalent training group or a control group. The attention abilities of the participants were assessed based on completing a continuous performance task that required response to specific target stimuli that were shown for only 250 ms. The results showed that the VR group had significantly higher rates of correct response and a decreased perceptual sensitivity and response bias when compared with the other two groups. Strickland, et al. [17] conducted two experimental studies to examine the effectiveness of VR therapy among children with autism spectrum disorder. The outcome was positive, i.e., learning through virtual reality is effective for children with autism, and the children engaged smoothly in the virtual environments as well as in following the instructions.

A study by Akhutina, et al. [18] examined the influence of the virtual environment on spatial functioning in children with cerebral palsy. The study compared 12-children with this condition who received VR therapy to a control group of 9-children with same condition who did not receive the VR therapy. A few of the children who received VR therapy showed some enhancement when compared with the other control group. Moreover, Denise Reid [19] examined the effectiveness of VR treatment for cerebral palsy in several studies. The focus of these studies was to investigate the effectiveness of VR treatment on feelings of self-efficacy, upper-extremity control, postural control and feelings of control and competency. In each of these studies, children diagnosed with different severities of cerebral palsy were involved in a series of interactive VR games. In three controlled studies by Reid [20], the effectiveness of this involvement on children's self-efficacy was examined based on interviews and performance measures. Results showed that VR treatment improved the self-efficacy of the children when compared to previous reports for the children.

The importance of the simulation used in VR or AR is to predict the capability, limitation and performance of a system and, as a result, will lead to a reduction in costs and hazards for experimentations compared with what is observed in traditional/hands-on methods [21]. An enhancement of those virtual/augmented simulated environments can be achieved through 106 agent-based modeling. The use of such a model produces a simulation with a high degree of fidelity [22]. In a study by Starner, et al. [23], the authors implemented an agent-based model in an augmented simulated environment using wearable computing devices and were able to develop a system that analyzes user activities and predict near-future possible needs. Another study, in [21], has proposed a new approach of introducing synthetic agents in motion at real-time into a real-life video depending on a terrain database and graphical rendering.

The rich literature on the confluence between augmented reality simulation and disorders' enhancement therapy has provided our approach with quite valuable inputs. The deployment of those components into this work in order to efficiently exploit their many advantages results in several features of the system that may include but not limited to intelligence, adaptiveness, responsiveness, and accuracy. This is because our interest is on developing a software system using an augmented reality simulated environment in order to help patients suffering from ADHD to find the treatment they need in the easiest and the fastest way possible, which has set it apart from traditional prior techniques.

## 3. AR-Therapist: Generic Design

This section defines the generic assembly of the model whereas the following section provide a case specific implementation and discussion.

*3.1. A Game Pipeline*

We present in this section a set of formal definitions to describe the logical flow for modeling the AR-Therapist.

1. **Treatment** is the whole treatment system (i.e., AR-Therapist). Its profile is a tuple of: ⟨Patient, Doctor, Game, {Treatment − Plan}⟩
    - Patient: Each patient will have his/her own profile. The profile has to be complete for the patient to join the treatment program.
    - Doctor: Each doctor will have his/her own profile.
    - Game: The game has to be defined by the doctor considering patients current mental state and the disorder severity.
    - {Treatment − Plan}: Players will go through a treatment program following a predefined.
    - treatment plan that includes playing an AR-based game consisting of a set of levels.
2. **Patient**: ⟨ID, Level, Performance − Index, {Preference}⟩
    - ID: is a short identification as a name or a referral number used by the doctor to define a patient.
    - Level: The level to where the patient has arrived in the treatment plan.
    - Performance − Index: The current value of performance the patient has achieved throughout the game.
    - {Preference}: The set of predefined preferences for a patient considering other psychological disorders that may affect current design and methodology of the treatment plan.
3. **Doctor**: ⟨ID, Experience, Involvement⟩
    - ID: The doctor has to have his/her own profile that is different from other therapists or psychological centers. This will give the doctor an access to the patient profiles and progress reports to allow for further evaluations.
    - Experience: Experience level of the doctor is useful in allowing access to more complex/detailed data of the patients.
    - Involvement: The level of engagement within the treatment process which allow the doctor to get involved in the game and in the reporting progress along the way of the patient assigned treatment.
4. **Game**: ⟨Type, {Level}⟩
    - Type: The type of the game to be played that has to be suitable for the patient. e.g., drag-and-drop and multiple-choices.
    - {Level}: The game consists of a set of levels that have different levels of complexity.
5. **Level**: ⟨{Object}, Max − Time, Effects⟩
    - {Object}: Maximum set of objects used in this level.
    - Max − Time: A predefined maximum time for the whole level to be completed or abort otherwise.
    - Effects: Simple directional voice or instructions used for guidance in case of a remote following.
6. **Object**: ⟨Shape, Size, Random − Location, Visibility⟩
    - Shape: The structure of an object has to be predefined beforehand the start of a session.
    - Size: The size of an object will depend on the location and closeness from the player focal point.
    - Random − Location: The initial distribution of objects around the real environment.
    - Visibility: The appearance of one object after another.
7. **Treatment-Program**: ⟨{Game − session}, Performance − Measures, Duration⟩
    - {Game − session}: The set of game session to complete the treatment program.
    - Performance − Measures: The performance in one session reports the correct, incorrect and uncompleted tried the patient has gone through in one session.
    - Duration: The maximum treatment time for the whole treatment program, e.g., 20 min to complete the treatment.
8. **Game-session**: ⟨Level, Timer, Current − Location, Number − of − Tries⟩

- Level: The game level has to be defined beforehand. The initial level is defined in the patient profile and player can move from one level to another asynchronously depending on his/her achievement in the session and then the patient profile is updated.
- Timer: To count the response time for the patients.
- Current − Location: To track current location of the patient for measuring closeness from objects within a session.
- Number − of − Tries: The repetition of tries within one session to include correct, incorrect and uncompleted tries. e.g., the number of collected target objects the patient has correctly collected in one session.

To this end, we have introduced eight-profiles that best formulate the AR-Therapist model when merged together. Next, we show the process of combining those profiles into a conceptual model.

*3.2. General Conceptual Model*

The conceptual model consists of four layers: (a) interface layer, (b) configuration layer, (c) run-time layer and (d) storage layer. Figure 1 depicts an architecture of the AR-Therapist based on the eight-profiles presented beforehand.

1. **The interface layer** contains the interfaces used for accessing other layers. For instance, doctor can use a user-friendly interface to configure the treatment plan, whereas players interact with the game through Augmented Reality glasses.
2. **The Configuration layer** consists of the "treatment plan configuration" component, where the doctor can add new treatment plan and configure the existence ones.
3. **The Run-time layer** has the components that interact with the player while the game is running. These components are: The Context-Agent component and the Game component.
    - **The Context Agent component** retrieves the player treatment plan, and his/her performance in order to control the current game-session and guide the player based on the treatment plan. Furthermore, the context agent capable of monitoring all the player's behaviors and interactions with the environment, logging "Ethically" the needed data, and calculating the player's performance following the equations highlighted in Section 3. Thus, the agent gains a deeper understanding of the patient to enrich the AR-Therapist model based on appropriate reasoning techniques. This in turns can be utilized in the future for suggesting the most optimal treatment plan for new patient (e.g., when facing a "cold start issue"). In addition, logging the needed data and explaining them will reveal some hidden information which may help the ADHD community.
    
        On the other hand, there is a persistent need for employing an intelligent agent capable of performing the tasks explained above since the AR-Therapist model is used by the patients with the aid of their families and under the doctors' remote supervision (i.e., ambient assistant living "AAL"). Surveying the literature resulted in finding prior use of AAL such as in fall detection [24,25], and for monitoring elderly in their homes using video surveillance [26,27].
    - **The game Component** involves levels of game; each level has its own games-sessions, and each game-session in turns contains the appropriate difficulty level that characterizes the maximum game-session time, used objects, and their locations.
4. **The storage layer** involves components which store data about the treatment-plan, treatment-program, player profile, and game levels and objects. Each component follows an appropriate common specification for structuring the related data. Thus, the system's extensibility can be guaranteed as well as the integration with the existence e-health systems.

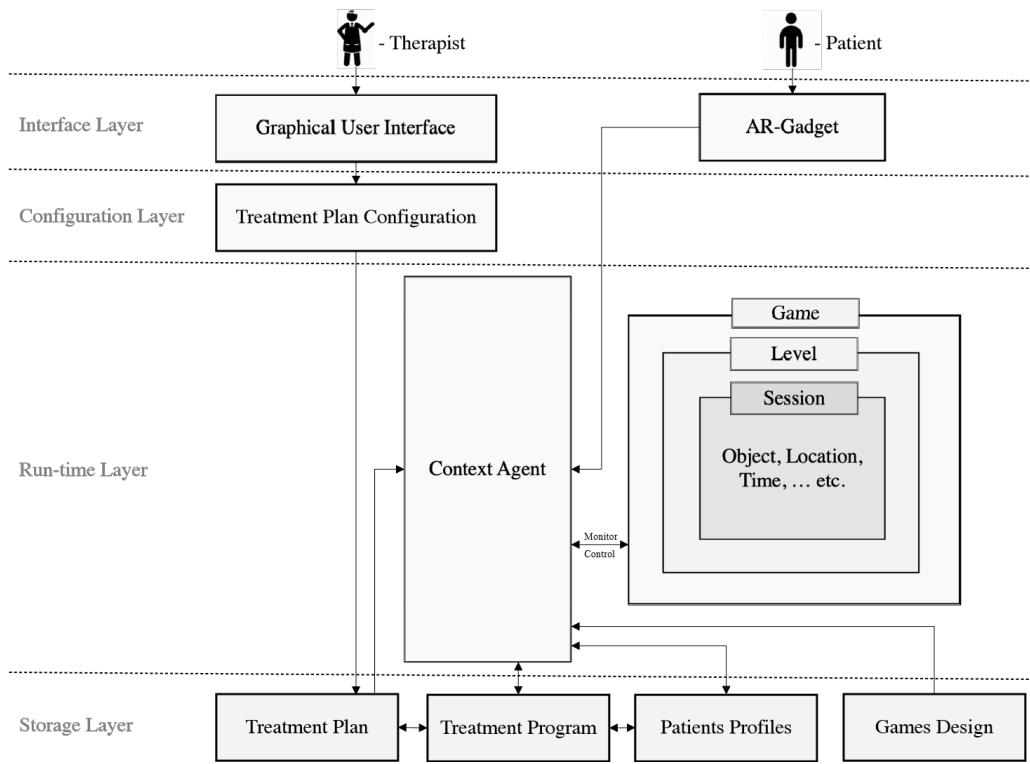

**Figure 1.** The general conceptual model for Augmented Reality (AR)-Therapist.

*3.3. Performance Measures*

One of the most important issues to diagnose and treat ADHD patients is to determine a set of accurate performance measures. These performance measures should have the ability to differentiate accurately between children having ADHD symptoms and others who do not have it. According to psychiatric recommendations, it would be better to collect these performance measures from children within different environments such as at home and at school [28].

There are many tests used to diagnose children with ADHD [29]. Continuous Performance Tests (CPT) are the most popular laboratory-based test supporting clinical diagnosis [30–32]. CPT is usually a computer-based test that aims to measure children attention and impulsivity. CPT involves the individual and random presentation of a series of visual or auditory stimuli that changes rapidly over a period of time. Children are informed to respond to the "target" stimulus and avoiding a "non-target" stimulus. The test provides summary statistics of performance parameters (e.g., response time, average response time, response time standard deviation, omission errors, and commission errors). These parameters have been shown to be useful in the detection of ADHD [33]. An important limitation of traditional CPTs is low ecological validity [34]. Ecological validity means the degree to which a performance test produces results similar to those produced in real life [35]. One approach to improve assessment methods which offers better ecological validity is CPTs based on VR, such as the Aula Nesplora test [36]. Those approaches have an advantage of being more realistic and ecologically-valid environment while still having the ability to assess the degree of ADHD severity. Using AR instead of VR will further improve the ecological validity of the performed test.

In this paper, multiple performance measures are used to provide ADHD diagnostics and treatment assessment. Thus, let us assume that we have an AR-game that frequently present an interactive environment with the following assumptions:

*T*: Number of tries in one session.
*C*: Number of correct tries in one session.
*I*: Number of incorrect tries (due to omission or commission errors) in one session.
*K*: Number of uncompleted tries in one session.

$$T = C + I + K \tag{1}$$

$$I = OE + CE \tag{2}$$

Then, the performance measures that will be used for providing ADHD diagnostics and treatment assessment include:

- Correct Response Times (*CRT*): The percentage of measuring attention deficits for the time spent on the correct tries.
- Mean of the CRT (*M*): To compare with correct response time to make sure they follow opposite relation to one another.

$$M = \sum_{i=1}^{C} \frac{CRT_i}{C} \tag{3}$$

- Standard deviation of the *CRT* (*SD*): Indicative of impulsive and hyperactive symptoms.

$$SD = \sqrt{\frac{1}{1-C} \cdot \sum_{i-1}^{C}(CRT_i - M)^2} \tag{4}$$

- Try time ($\theta$): The maximum allowed time to complete one try within a session.
- Omission errors (*OE*): The absence of any response during a try period to be used to measure inattention.
- Commission errors (*CE*): The response to non-target stimuli which to be used to measure impulsivity.
- Engagement Factor (*GF*): It indicates the patient engagement level with the game.

$$EF = \frac{C + I}{T} \tag{5}$$

- Inattention Factor (*IAF*): It represents the percentage of patient's inattention.

$$IAF = \frac{OE}{C + I} \tag{6}$$

- Impulsivity Factor (*IMF*): Indicative of percentage of the patient's impulsivity observed in his/her behavior within a session.

$$IMF = \frac{CE}{C + I} \tag{7}$$

- Error Factor (*EF*): It represents the percentage of the error rate during a session.

$$EF = \frac{OE + CE}{C + I} \tag{8}$$

- Correct Response Factor (*CRF*): The percentage of the total correct response time relatively to maximum allowed time for all correct tries.

$$CRF = \frac{\sum_{i=1}^{C} CRT_i}{C \times \theta} \tag{9}$$

- Performance Index (*PI*): It reflects the single measure for the overall performance of the patient which depends on the correct response factor, error factor, and engagement factor.

$$PI = \left[\frac{(1 - CRF) + (1 - EF)}{2}\right] \times GF \quad (10)$$

Previous measurements are normalized within the interval (0–1). In Equation (3), *CRT* measures the length of time that the child takes to make a correct try (i.e., choose the target object in the try). The longer the *CRT* is, the more likely the child has attention deficit. This is because one of the symptoms of attention deficit is that the child cannot focus on tasks. As a result, the child takes a longer time compared to normal children when doing a task (i.e., choosing the target object in our case). We use the mean of all *CRT*s (*M*) in the game session to measure the attention deficit of the child. In addition, the standard deviation of *CRT*s (*SD*) is used in Equation (4) to indicate the impulsivity and hyperactivity of the child. The higher the value of *SD*, the more probability that the child suffers from impulsivity and hyperactivity. A child with impulsivity and hyperactivity has difficulty in controlling his moves after a certain period of time. As a result, the child starts periodically to move with no destination. Such a child in our case, will have great differences among *CRT*s because the impulsivity and hyperactivity will hinder him from moving towards the target object in some tries. Engagement Factor (*GF*), in Equation (5), indicates the engagement-level of the child in the game. In our case, the child is considered to be engaged in the game of s/he keeps playing the game. In contrast, the child is considered to be not engaged if s/he stops the game before completing all tries in the session. Thus, *GF* is defined as the number of correct and incorrect tries (*C* + *I*) divided by the total number of tries (*T*) in the session.

Inattention Factor (*IAF*) which is defined in Equation (6) indicates the child inattention. In our case, the child inattention increases when he makes Omission Errors (*OE*) in the session, i.e., when the patient does not choose any of the objects appearing to him/her. The number of uncompleted tries (*K*) in the session should be excluded when indicating *IAF*. Thus, *IAF* is defined as the number of *OE*s divided by $C + (I \cdot C) + I = T - K$. Impulsivity Factor (*IMF*) in Equations (7) is defined as the number commission errors divided by number of correct and incorrect tries. We also exclude *K* when defining *IMF*. In our case, the child who suffers from impulsivity will likely make more commission errors because impulsivity will prevent him/her from focusing when choosing an object. Error Factor (*EF*), given in Equation (8), indicates the child error rate. The error in our case includes omission and commission errors excluding *K*. Thus, *EF* is equal to $IAF + IMF$. Correct Response Factor (*CRF*), in Equation (9), measures the percentage of the correct response of the child in one session. In our case, *CRF* should be negatively affected by the amount of time that the child takes when he makes an incorrect try. Thus, we define *CRF* as the total summation of *CRT*s to the actual time of the game during the session (*GT*). In this case, *CRF* will be 100% if the child makes all tries correctly. Otherwise, it will decrease depending on the total amount of time spent by the child on incorrect tries.

The final performance measure which is given in Equation (8) is the Performance Index (*PI*). *PI* is a composite score which measures the overall performance of the child. In our case, we want the *PI* of the child to be affected positively by his *CRF* and negatively by his *EF*. In addition, we need to take into account different possible scenarios that can happen in the game session. One possible scenario is that the child does not complete all the tries in the session. The child can make one correct try and stop the game before finishing all the tries in the session. If we only considered the *CRF* and *EF*, the *PI* in the case would be the highest. In order to prevent this from happening, we consider the *GF* in the definition of *PI*. Another possible scenario is that we have two children who have the same *CRF*, *EF*, and *GF* but different *GT*. In this case, they will have the same *PI*. However, the child who has less *GT* should have a higher *PI*. Thus, we considered the ratio of *GT* to the maximum Session Time (*ST*) in the definition of *PI*. The *PI* of the child should be affected negatively by this ratio.

## 4. Implementation

Patients with a predefined ADHD disorder are having difficulty concentrating on a specific task. Therapists usually use simple concentration games with children in order to help with increasing their concentration. We imitate the usefulness of traditional games using an augmented reality environment. The use of the AR environment is more suitable to children than virtual ones since it allows them to observe their current location while the game is being introduced. Major benefit of the game-based environment entails on-time measurements and instant reporting which in part makes the job of specialized therapists much easier.

The game is simply a simulation of two 3D-balls: one is the target ball, and another is not. The player should follow the instructions and hits the target ball with a specified time period which will add to the values of correct hits. If not, the value will be added to the incorrect hits as an omission or commission error and taking out from the total number of tries the player has to complete to finish a session. We used Unity in the design of the AR game, and the test was through the Microsoft HoloLens emulator. Figure 2 shows a screenshot of the game before and after the patient hits the target ball.

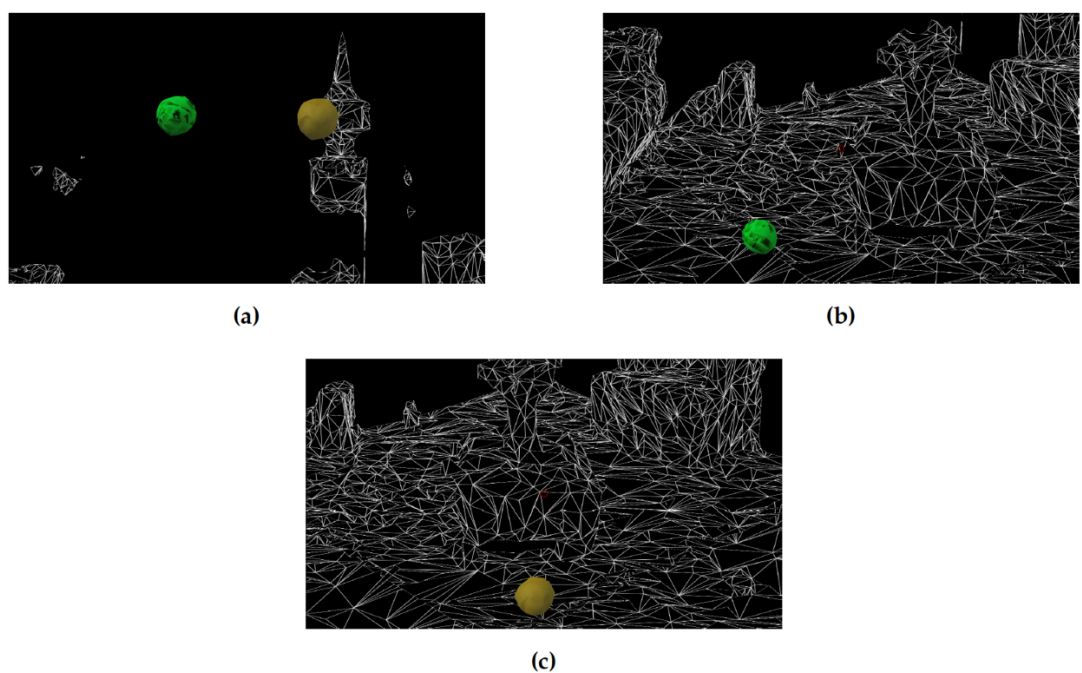

**Figure 2.** Basic design of the game where one ball is the target while another is the non-target. Above images are before and after hitting the target ball or the wrong ball: (**a**) Initial screen, (**b**) Dropping the correct ball, and (**c**) Dropping the wrong ball.

The child can play many game sessions during one treatment session. The number of sessions depend on his performance in the game and his engagement level. Currently, the game consists of only one level. Each game session consists of 10-trials of one minute each. In each trial, the "target" and the "non-target" ball will appear to the child. The child will be directed to catch the target ball so that it will be dropped down during the trial interval. The time elapsed from the beginning of the trial until hitting the target is the correct response time. If this time decreases, this is an indicator that the performance of the child has improved. If the child was not able to catch any of the balls, this is an omission error which indicates inattention. If s/he caught the non-target ball, this is a commission error which indicates impulsivity. The proposed game-based model can be described using the following steps:

1. The child starts the game session.

2. A brief explanation of the game is presented to the child along with an indication of the target ball.
3. A timer is started to count 60 seconds of the first trial.
4. The trial is ended in four situations. Firstly, if the child selected the target ball. Secondly, if s/he selected the non-target ball (commission error). Thirdly, the time is out, and s/he did not choose any ball (omission error). Finally, if s/he ends the game.
5. The trial statistical data are recorded.
6. If it is not the tenth trial or the game is not ended, the next trial begins.
7. All performance parameters are calculated and recorded for further analysis.

## 5. Experimentation with a Case Study

A case study is developed as a simulation of practicing the game. The purpose of the case study is to validate the accuracy and effectiveness of the proposed model. The case study is divided into two parts. In the first part, we need to study the effect of number of correct tries and number of errors on the performance index. While in the second part, we need to study the effect of patient's engagement in the game on the performance index.

The first part of the experiment consists of three phases, each phase uses 20 sessions and represents different combination of the number of correct tries ($C$), the number of omission errors ($OE$), the number of Commission errors ($CE$), and the number of uncompleted tries ($K$). In all phases, we consider the value of $K = 0$. This is due to the need to show only the effect of the number of correct tries versus the number of errors on the value of the performance index ($PI$). In the first phase, we set the value of $C = 3$ and $I = 7$. In the second phase, we set the value of $C = I = 5$. Finally, in the third phase, we set the value of $C = 8$ and $I = 2$. Table 1 shows the values of different parameters and performance measures used in the first part of the experiment. In Table 1, we assumed that the patient was fully engaged in the game during the three phases ($K = 0$ and $GF = 100\%$). We can also notice that the mean ($M$) and standard deviation ($SD$) values of the correct response times are very similar across all three phases. In each successive phase, the $PI$ values continue to improve depending only on the number of correct and error tries.

**Table 1.** Summary of statistical data obtained during the first part of the experiment.

| Experiment Parameters & Performance Measures | Phase 1 | Phase 2 | Phase 3 |
|---|---|---|---|
| No. of Sessions | 20 | 20 | 20 |
| $C$ | 3 | 5 | 8 |
| $OE$ | 3 | 3 | 1 |
| $CE$ | 4 | 2 | 1 |
| $K$ | 0 | 0 | 0 |
| $IAF$ | 30% | 30% | 10% |
| $IMF$ | 40% | 20% | 10% |
| $EF$ | 70% | 50% | 20% |
| $GF$ | 100% | 100% | 100% |
| $M$ | 25.13 | 23.71 | 21.81 |
| $SD$ | 14.46 | 14.17 | 13.58 |
| Average $PI$ | 44% | 55% | 72% |

Figure 3 shows three $PI$ curves corresponding to the three phases of the experiment. In each phase, we assumed that the correct response time for each patient to improve gradually with each trial in each session and in successive sessions as the patient continues to get more familiar with game and its environment. The $PI$ values improve during all sessions in the same phase of the experiment due to the improvement in the correct response time from one session to the next. Also, we can observe that the *PI* values improve further from one phase to the next as the number of errors decreased and the number of correct tries increased.

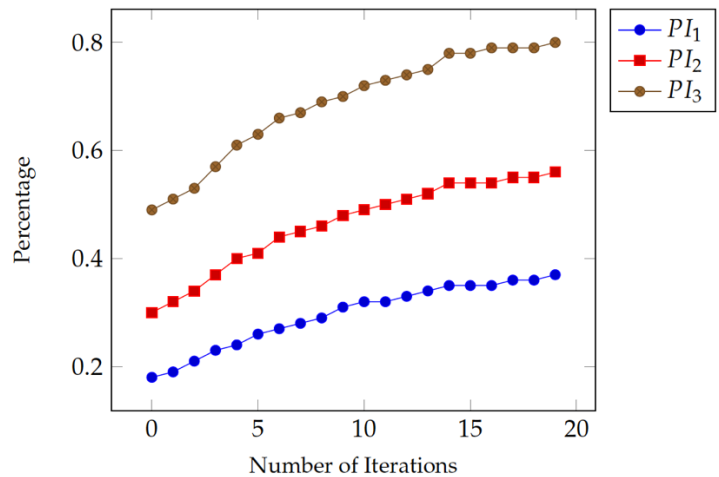

**Figure 3.** Different values for the performance index, by implementing Equation (9). The lower line presents a static value of *C* = 3 and *I* = 7. The mid-line presents an equal value for *C* = *I* = 5. The upper line presents the values of *C* = 8 while *I* = 2.

The second part of the experiment consists of four phases, each phase also uses 20 sessions. For all phases, the number of errors is considered to be zero (*EF* = 0%). Only the effect of changing the numbers of correct tries and uncompleted tries (patient engagement) are considered. In the first phase, we set the value of *C* = 3 and *K* = 7. In the second phase, we set the value of *C* = 5 and *K* = 5. In the third phase, we set the value of *C* = 7 and *K* = 3. Finally, in the fourth phase, we set the value *C* = 10 and *K* = 0.

Table 2 shows the values of different parameters and performance measures used in the second part of the experiment. The assumption is that the patient did not make any errors during the four phases (*OE* = 0, *CE* = 0, and *EF* = 0%). Here again, we can notice that the mean (*M*) and standard deviation (*SD*) values of the correct response times are very similar across all four phases. In each successive phase, the *PI* value progresses based only on the number of completed tries during the game sessions (*GFs*).

**Table 2.** Summary of statistical data obtained during the second part of the experiment.

| Experiment Parameters & Performance Measures | Phase 1 | Phase 2 | Phase 3 | Phase 4 |
|---|---|---|---|---|
| No. of Sessions | 20 | 20 | 20 | 20 |
| *C* | 3 | 5 | 7 | 10 |
| *OE* | 0 | 0 | 0 | 0 |
| *CE* | 0 | 0 | 0 | 0 |
| *K* | 7 | 5 | 3 | 0 |
| *IAF* | 0% | 0% | 0% | 0% |
| *IMF* | 0% | 0% | 0% | 0% |
| *EF* | 0% | 0% | 0% | 0% |
| *GF* | 30% | 50% | 70% | 100% |
| *M* | 25.13 | 23.93 | 22.94 | 21.62 |
| *SD* | 14.46 | 14.13 | 13.80 | 13.18 |
| Average *PI* | 24% | 40% | 57% | 82% |

In each phase, we assumed that the correct response time of the patient improved gradually with each trial in each session and also in successive sessions as the patient gets more familiar with the game and its environment.

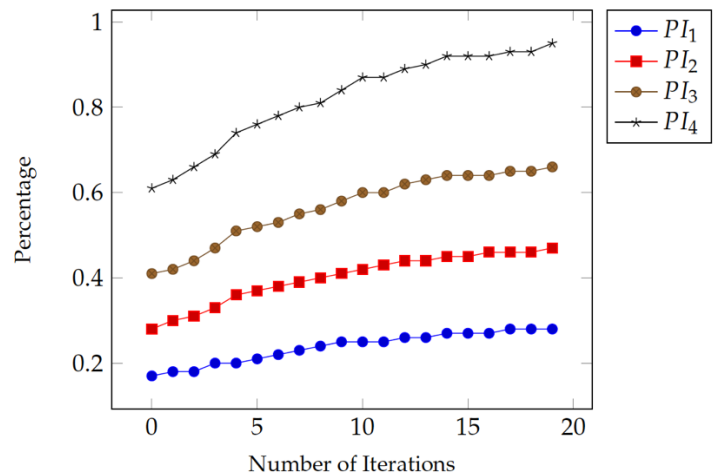

**Figure 4.** Different values for the performance index, by implementing Equation (9). The lower line presents a static value of *C* = 3 and *K* = 7. The second line present an equal value for *C* = *K* = 5. The third line presents a static value of *C* = 7 and *K* = 3. The upper line presents the values of *C* = 10 while *K* = 0.

As Figure 4 shows, the *PI* values are getting improved during all sessions in the same phase of the experiment due to the improvement in the correct response time from one session to the next. Also, we can observe that the *PI* values are further improved from one phase to the next as the number of completed tries, i.e., patient engagement, increases and the number of uncompleted tries decreases.

## 6. Conclusion and Future Work

The paper proposed a theoretical cognitive model that helps to enhance behavior of patients with a predefined ADHD using a game-based augmented reality environment that we called "AR-Therapist". The purpose is to provide an online alternative to traditional CBT with a more advanced virtual one that may exceed traditional CBT methods with higher time and resource efficiency. The model has been implemented on a simulated augmented reality environment as a simple drop-the-ball game. The results have shown that there is an increase in patient's correct attention to choose a predefined object which contributes positively to their performance index, i.e., they are following along with their treatment plan. AR-Therapist achieves an excellent accessibility level to every patient in need as it mimics the therapist role through utilizing an augmented reality game-based environment to allow for features such as: adaptiveness, smartness, responsiveness, and accuracy. Other advantages are availability and assurance of the therapist's level-of-experience which cannot be guaranteed in traditional CBT. Future work will provide evaluations involving human subjects in order to compare the proposed AR-Therapist with traditional CBT.

**Author Contributions:** Conceptualization, S.A., M.A. and A.A.; formal analysis, S.A. and M.A.; funding acquisition, S.A.; investigation, A.M.; methodology, S.A., A.A. and A.M.; project administration, S.A.; software, A.M.; supervision, S.A.; evaluation, M.A.; visualization, S.A. and A.M.; writing–original draft, S.A.; writing–review & editing, S.A. and A.M.

**Funding:** This work was funded by the Deanship of Scientific Research at Albaha University, Saudi Arabia (grant number: 1439/4). Any opinions, findings, and conclusions or recommendations expressed in this material are those of the author(s) and do not necessarily reflect the views of the Deanship of Scientific Research or Albaha University.

**Conflicts of Interest:** The authors declare no conflict of interest.

## References

1. Alqithami, S.; Alzahrani, M.; Alzahrani, A.; Ahmed, Y. Modeling an Augmented Reality Game Environment to Enhance Behavior of ADHD Patients. The 12th International Conference on Brain Informatics. *Springer* **2019**, in press.
2. Barkley, R.A. The ecological validity of laboratory and analogue assessment methods of ADHD symptoms. *J. Abnorm. Child Psychol.* **1991**, *19*, 149–178.
3. Abikoff, H.; Courtney, M.; Pelham,W.E.; Koplewicz, H.S. Teachers' ratings of disruptive behaviors: The influence of halo effects. *J. Abnorm. Child Psychol.* **1993**, *21*, 519–533.
4. Parsons, T.D.; Bowerly, T.; Buckwalter, J.G.; Rizzo, A.A. A controlled clinical comparison of attention performance in children with ADHD in a virtual reality classroom compared to standard neuropsychological methods. *Child Neuropsychol.* **2007**, *13*, 363–381.
5. Gorini, A.; Gaggioli, A.; Vigna, C.; Riva, G. A second life for eHealth: prospects for the use of 3-D virtual worlds in clinical psychology. *J. Med. Int. Res.* **2008**, *10*, e21.
6. Beard, L.;Wilson, K.; Morra, D.; Keelan, J. A survey of health-related activities on second life. *J. Med. Int. Res.* **2009**, *11*, e17.
7. Parsons, T.D.; Rizzo, A.A.; Rogers, S.; York, P. Virtual reality in paediatric rehabilitation: a review. *Dev. Neurorehabilit.* **2009**, *12*, 224–238.
8. Bickmore, T.W.; Mitchell, S.E.; Jack, B.W.; Paasche-Orlow, M.K.; Pfeifer, L.M.; O'Donnell, J. Response to a relational agent by hospital patients with depressive symptoms. *Interact. Comput.* **2010**, *22*, 289–298.
9. Meyerbröker, K.; Emmelkamp, P.M. Virtual reality exposure therapy in anxiety disorders: a systematic review of process-and-outcome studies. *Depress. Anxiety* **2010**, *27*, 933–944.
10. Azuma, R.T. A Survey of Augmented Reality. Presence: *Teleoperators Virtual Environ.* **1997**, *6*, 355–385. doi:10.1162/pres.1997.6.4.355.
11. Burdea, G.C.; Coiffet, P. Virtual Reality Technology; *John Wiley & Sons*, 2003.
12. Billinghurst, M.; Clark, A.; Lee, G. A Survey of Augmented Reality. *Found. Trends Hum. Comput. Interact.* **2015**, *8*, 73–272.
13. Biederman, J. Attention-deficit/hyperactivity disorder: a selective overview. *Biol. Psychiatry* **2005**, *57*, 1215–1220.
14. Schachar, R.; Mota, V.L.; Logan, G.D.; Tannock, R.; Klim, P. Confirmation of an inhibitory control deficit in attention-deficit/hyperactivity disorder. *J. Abnorm. Child Psychol.* **2000**, *28*, 227–235.
15. Ben-Moussa, M.; Rubo, M.; Debracque, C.; Lange,W.G. Djinni: a novel technology supported exposure therapy paradigm for sad combining virtual reality and augmented reality. *Front. Psychiatry* **2017**, *8*, 26.
16. Cho, B.H.; Ku, J.; Jang, D.P.; Kim, S.; Lee, Y.H.; Kim, I.Y.; Lee, J.H.; Kim, S.I. The effect of virtual reality cognitive training for attention enhancement. *Cyberpsychol. Behav.* **2002**, *5*, 129–137.
17. Strickland, D.; Marcus, L.M.; Mesibov, G.B.; Hogan, K. Brief report: Two case studies using virtual reality as a learning tool for autistic children. *J. Autism Dev. Disord.* **1996**, *26*, 651–659.
18. Akhutina, T.; Foreman, N.; Krichevets, A.; Matikka, L.; Narhi, V.; Pylaeva, N.; Vahakuopus, J. Improving spatial functioning in childrenwith cerebral palsy using computerized and traditional game tasks. *Disabil. Rehabil.* **2003**, *25*, 1361–1371.
19. Reid, D. Changes in seated postural control in children with cerebral palsy following a virtual play environment intervention: a pilot study. *Isr. J. Occup.* **2002**, *11*, E75–E95.
20. Reid, D.T. Benefits of a virtual play rehabilitation environment for children with cerebral palsy on perceptions of self-efficacy: a pilot study. *Pediatric Rehabil.* **2002**, *5*, 141–148.
21. Gelenbe, E.; Hussain, K.; Kaptan, V. Simulating autonomous agents in augmented reality. *J. Syst. Softw.* **2005**, *74*, 255–268.
22. Shendarkar, A.; Vasudevan, K.; Lee, S.; Son, Y.J. Crowd simulation for emergency response using BDI agents based on immersive virtual reality. *Simul. Model. Pract. Theory* **2008**, *16*, 1415–1429.
23. Starner, T.; Mann, S.; Rhodes, B.; Levine, J.; Healey, J.; Kirsch, D.; Picard, R.W.; Pentland, A. Augmented reality through wearable computing. *Presence: Teleoperators Virtual Environ.* **1997**, *6*, 386–398.

24. Zhang, T.;Wang, J.; Liu, P.; Hou, J. Fall detection by embedding an accelerometer in cellphone and using KFD algorithm. *IJCSNS* **2006**, *6*(10), 277–284.
25. Chan, E.;Wang, D.; Pasquier, M. Towards intelligent self-care: Multi-sensor monitoring and neuro-fuzzy behavior modelling. *2008 IEEE Int. Conf. Syst. Man Cybern.* **2008**, pp. 3083–3088. doi:10.1109/ICSMC.2008.4811769.
26. Nasution, A.H.; Emmanuel, S. Intelligent Video Surveillance for Monitoring Elderly in Home Environments. *2007 IEEE 9th Workshop Multimed. Signal Process.* **2007**, pp. 203–206. doi:10.1109/MMSP.**2007**.4412853.
27. Foroughi, H.; Aski, B.S.; Pourreza, H. Intelligent video surveillance for monitoring fall detection of elderly in home environments. *2008 11th Int. Conf. Comput. Inf. Technol.* **2008**, pp. 219–224. doi:10.1109/ICCITECHN.2008.4803020.
28. Morales-Hidalgo, P.; Hernández-Martínez, C.; Vera, M.; Voltas, N.; Canals, J. Psychometric properties of the Conners-3 and Conners Early Childhood Indexes in a Spanish school population. *Int. J. Clin. Health Psychol.* **2017**, *17*, 85–96.
29. Hall, C.L.; Valentine, A.Z.; Groom, M.J.; Walker, G.M.; Sayal, K.; Daley, D.; Hollis, C. The clinical utility of the continuous performance test and objective measures of activity for diagnosing and monitoring ADHD in children: a systematic review. *Eur. Child Adolesc. Psychiatry* **2016**, *25*, 677–699.
30. Edwards, M.C.; Gardner, E.S.; Chelonis, J.J.; Schulz, E.G.; Flake, R.A.; Diaz, P.F. Estimates of the validity and utility of the Conners' Continuous Performance Test in the assessment of inattentive and/or hyperactive-impulsive behaviors in children. *J. Abnorm. Child Psychol.* **2007**, *35*, 393–404.
31. Vogt, C.; Williams, T. Early identification of stimulant treatment responders, partial responders and non-responders using objective measures in children and adolescents with hyperkinetic disorder. *Child Adolesc. Ment. Health* **2011**, *16*, 144–149.
32. Conners, C.K.; Staff, M.; Connelly, V.; Campbell, S.; MacLean, M.; Barnes, J. Conners' continuous performance Test II (CPT II v. 5). *Multi-Health Syst. Inc.* **2000**, *29*, 175–196.
33. Rapport, M.D.; Chung, K.M.; Shore, G.; Denney, C.B.; Isaacs, P. Upgrading the science and technology of assessment and diagnosis: Laboratory and clinic-based assessment of children with ADHD. *J. Clin. Child Psychol.* **2000**, *29*, 555–568.
34. Rodríguez, C.; Areces, D.; García, T.; Cueli, M.; González-Castro, P. Comparison between two continuous performance tests for identifying ADHD: Traditional vs. virtual reality. *Int. J. Clin. Health Psychol.* **2018**, *18*, 254–263.
35. Negut, A.; Jurma, A.M.; David, D. Virtual-reality-based attention assessment of ADHD: ClinicaVR: Classroom-CPT versus a traditional continuous performance test. *Child Neuropsychol.* **2017**, *23*, 692–712.
36. Climent, G.; Banterla, F.; Iriarte, Y. AULA: Theoretical manual. *San Sebastianspain* **2011**.